\documentclass[
onecolumn,
tightenlines,
11pt,
longbibliography,
showpacs,
nofootinbib,
notitlepage,
superscriptaddress]{revtex4-2}

\usepackage[utf8]{inputenc} 
\usepackage[T1]{fontenc}    
\usepackage{hyperref}       
\usepackage{url}            
\usepackage{booktabs}       
\usepackage{amsfonts}       
\usepackage{nicefrac}       
\usepackage{microtype}      
\usepackage{xcolor}         

\usepackage{braket}
\usepackage{amssymb}
\usepackage{amsmath}
\usepackage[capitalize, nameinlink]{cleveref}
\makeatletter
\AddToHook{cmd/appendix/before}{\def\cref@section@alias{appendix}}
\makeatother
\usepackage{subcaption}
\usepackage{float}
\usepackage{tikz}
\usepackage{algpseudocode}
\usetikzlibrary{arrows.meta, positioning, shapes.geometric, fit}

\usepackage{caption}
\hypersetup{
  colorlinks   = true, 
  urlcolor     = blue, 
  linkcolor    = red, 
  citecolor   = blue 
}

\begin{document}

\title{Diffusion-warm sampling of the XY model \\ enables fast thermalization at scale}

\author{Sehmimul Hoque}
\email{s4hoque@uwaterloo.ca}
\affiliation{Institute for Quantum Computing, University of Waterloo, Waterloo, Ontario N2L 3G1, Canada}
\affiliation{Department of Physics and Astronomy, University of Waterloo, Waterloo, Ontario N2L 3G1, Canada}
\affiliation{Perimeter Institute for Theoretical Physics, Waterloo, Ontario N2L 2Y5, Canada}

\author{Roger Melko}
\email{rmelko@perimeterinstitute.ca}
\affiliation{Department of Physics and Astronomy, University of Waterloo, Waterloo, Ontario N2L 3G1, Canada}
\affiliation{Perimeter Institute for Theoretical Physics, Waterloo, Ontario N2L 2Y5, Canada}

\author{Pooya Ronagh}
\email{pooya.ronagh@uwaterloo.ca}
\affiliation{Institute for Quantum Computing, University of Waterloo, Waterloo, Ontario N2L 3G1, Canada}
\affiliation{Department of Physics and Astronomy, University of Waterloo, Waterloo, Ontario N2L 3G1, Canada}
\affiliation{Perimeter Institute for Theoretical Physics, Waterloo, Ontario N2L 2Y5, Canada}
\affiliation{Microsoft, Redmond, WA 98052, USA}

\begin{abstract}
We introduce a novel technique for scalable sampling of spin-system states with continuous symmetries using diffusion models.
By applying our approach to the XY model, a fundamental continuous-spin model in condensed matter physics, we show that our technique addresses the shortfalls of the Markov chain Monte Carlo (MCMC) in generalization to varying system sizes. More specifically, we show that training a temperature-conditioned diffusion model on smaller-size XY model lattices enables the generation of accurate samples in larger lattice sizes. By tracking physically important observables of the model, such as spin correlations, our experiments demonstrate that diffusion sampling followed by a few MCMC steps reduces the thermalization time by an order of magnitude relative to the standard MCMC with random initialization. Our study provides valuable insight as to how generative models can be used to study continuous-state condensed matter systems at scale.
\end{abstract}
\maketitle

\section{Introduction}
Physical systems often exhibit continuous degrees of freedom. This includes lattice models in condensed matter physics, gauge theories, molecular conformational searching, and many other optimization settings involving continuous spins, fractional current and voltage values, etc. So far, computational physicists have developed many efficient algorithms for simulating equilibrium and non-equilibrium dynamics of physical systems with discrete state spaces (such as the Ising model \cite{Ising1925}, Potts models \cite{Potts_1952}, clock model \cite{clock-model}, etc.). However, these methods struggle to explore continuous-variable landscapes effectively. Moreover, in many such systems the continuous degrees of freedom have a periodic nature, which further confuses exploration strategies relying on local information and local moves, such as gradient-based or other convex optimization techniques. 

The 2D XY model is a fundamental spin model in condensed matter physics where the spins can take continuous values between $0$ to $2\pi$ on a two dimensional lattice. It exhibits a Berezinskii-Kosterlitz-Thouless phase transition and has been studied extensively using numerical and analytical tools \cite{bkt_1973, xy_critical_temp}. The XY model is a fundamental model of magnetization in devices and materials, as well as superfluids and superconducting systems \cite{josephson-junction, superconducting-films}. Additionally the XY model has been associated with complex valued neural networks, implying that efficient sampling from XY model can have potential applications in neural network training \cite{complex-neural, xy-laser}. The anticipated high impact of solving the XY model efficiently has even led to the advent of special-purpose computers for simulating them using various species of coherent bosonic networks \cite{xy-laser, photonic-ising-machine, non-degenerate}.

In this paper, we focus on simulating the equilibrium dynamics of the XY model. A central yet computationally intractable problem in this setting is Gibbs sampling, wherein, given an energy function $E(x)$ corresponding to system configurations $x$, and an ambient temperature $T>0$, the task is to generate high-quality independent samples from the distribution $p(x) = e^{-E(x)/T}/\mathcal{Z}$ where the normalization constant $\mathcal{Z}$ is the \emph{partition function}. The standard approach to generate such independent Gibbs samples from the XY model is Markov Chain Monte Carlo using Wolff steps \cite{mcmc-revolution, wolff}. However, MCMC can struggle with very long thermalization times, defined as the time taken for the Markov Chain to evolve from its initial random state to its stationary equilibrium distribution \cite{glauber_mixing, stat-mech-book}.

\begin{figure}[t]
    \centering
    \begin{subfigure}{0.495\linewidth}
        \centering
        \includegraphics[width=\linewidth]{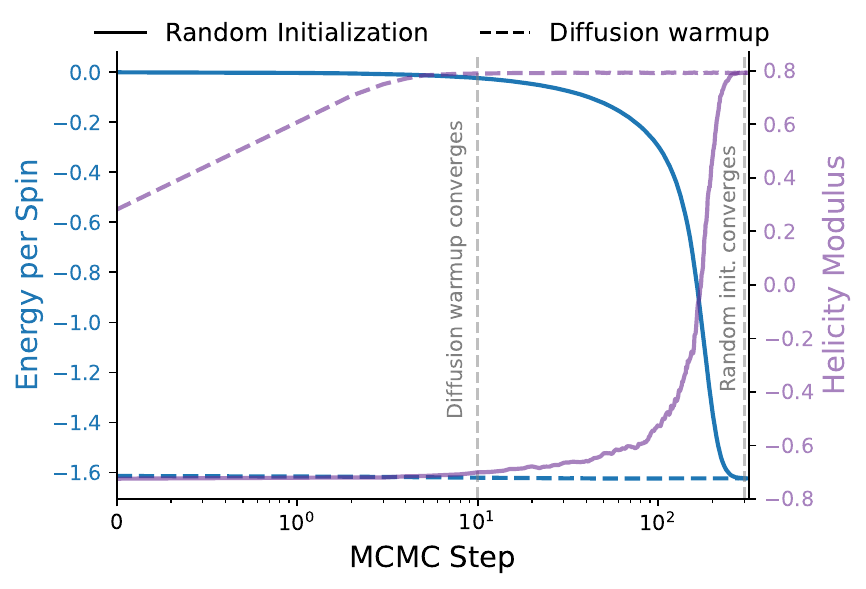}
        \label{fig:energy-helicity-dual}
    \end{subfigure}
    \hfill
    \begin{subfigure}{0.495\linewidth}
        \centering
        \includegraphics[width=\linewidth]{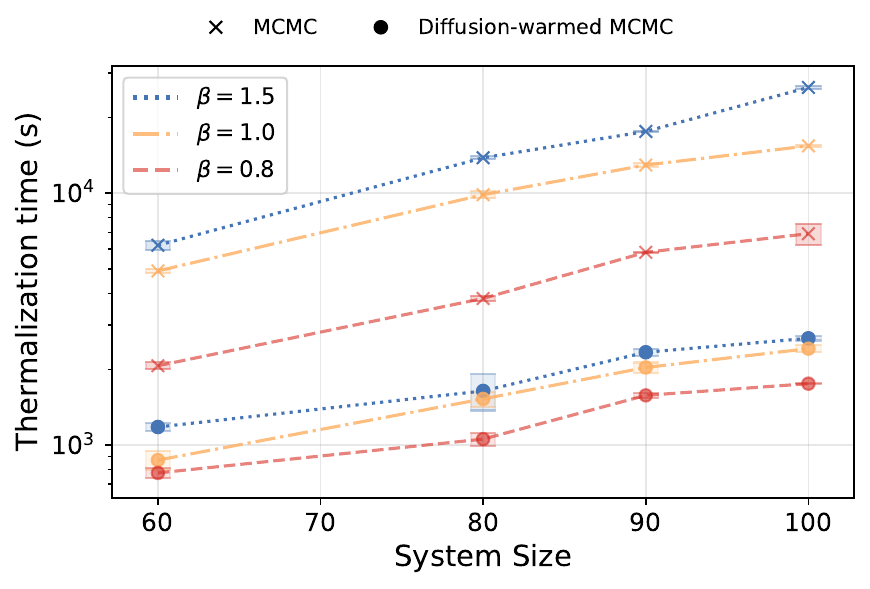}
        \label{fig:runtime}
    \end{subfigure}
    \caption{The figure demonstrates samples generated from the diffusion model provide an efficient warm start for MCMC, substantially reducing the thermalization time required to reach equilibrium configurations in large XY model systems. The diffusion model was trained on $20\times 20$ XY model samples on a grid of temperatures ranging from $\beta = 0.5$ to $\beta = 1.6$. \textbf{Left panel:} Energy per spin (blue) and helicity modulus (purple) compared against MCMC steps for randomly initialized samples (solid lines) versus samples initialized from the trained diffusion model (dashed lines) for a $60\times60$ system at $\beta = 1.5$. The system initialized using the diffusion samples thermalizes in very few MCMC steps, demonstrating that the diffusion samples are a very good warm start for MCMC. The dashed vertical lines show the steps when the observables initialized in the two different ways converge. \textbf{Right panel:} Total runtime for thermalization at different temperatures across the phase transition plotted against system size for pure MCMC and the diffusion-warmed MCMC method. The shaded area represents the standard deviation error bars in thermalization time. In the diffusion-warmed MCMC method, the trained diffusion model is used to do inference starting from random configurations and then a few steps of MCMC are done until thermalization (see \cref{app:thermalization} for details on computing thermalization time). This demonstrates approximately one order of magnitude speedup in thermalization time using diffusion-warmed MCMC method compared to the pure MCMC method.}
    \label{fig:main-result}
\end{figure}
Machine learning has been explored for Gibbs sampling of the XY model to accelerate sampling by training neural networks to approximate the target Boltzmann distribution \cite{flow-xy}. Models trained on data from fixed or multiple lattice sizes can generate accurate samples within the training regime, but often struggle to generalize to larger system sizes \cite{rydberg-gpt}. In particular, extrapolation beyond the training size can lead to exponentially growing errors \cite{flow-xy}, which limits reliable Gibbs sampling for larger systems. Another approach to generate Gibbs samples from the XY model is by variational training where a model is trained without data. Variational training is generally done by importance sampling where samples are generated by the model during different steps of training and then weights are assigned depending on the probability of the sample and the Gibbs density \cite{importance-sampling-size, boltzmann-gen}. However, these models suffer from mode collapse in multimodal and high dimensional systems where some modes dominate the importance weights and thus causing low probability regions to be under represented \cite{mode-collapse}. Furthermore, the density of the samples produced by the model is also not readily available for some models and thus requires estimation of the density which can be computationally expensive. This can make training such models unstable and computationally very expensive. 

In this work we propose a different approach to obtain accurate samples from the Gibbs distribution for different system sizes overcoming both of the challenges stated above. We show that by training a rotationally invariant conditional diffusion model on a small $20\times 20$ lattice system, we can perform inference at larger systems with small number of MCMC steps to obtain accurate samples. We call this method diffusion-warmed MCMC. Our main result is that diffusion-warmed MCMC reduces the thermalization time by around an order of magnitude compared to standard MCMC methods across all the system sizes and temperatures we have studied. In \cref{fig:main-result} we demonstrate this reduction in thermalization time across different temperatures and systems sizes using the diffusion-warmed MCMC method versus pure MCMC. In both cases, we start for $1000$ randomly initialized configurations.

We also investigate why a naive extrapolation in system size without any MCMC steps does not allow us to produce accurate samples from the XY model. Additionally, we compare our technique against a tiling baseline that constructs larger lattices by repeating small-lattice samples. Our experiments show that tiling underestimates long-range correlations and introduces artificial artifacts (more discussion on this is given in \cref{app:tiling}).

Our diffusion-warmed MCMC method introduces a novel way to accelerate equilibration in XY models by leveraging the learned dynamics from diffusion models. In \cref{sec:background} we first give some background on the condensed matter system studied here and then explain our method in \cref{sec:method}. In \cref{sec:results}, we demonstrate the necessity of incorporating MCMC steps to obtain reliable samples and show a clear reduction in the time required to reach equilibrium compared to traditional MCMC. We also evaluate the accuracy of our method on larger lattices by comparing several observables with ground-truth samples. In \cref{sec:conclusion} we discuss the limitations of diffusion-warmed MCMC and outline several directions for future work.

\section{Background}
\label{sec:background}
In this section we first discuss some aspects of the two dimensional XY model and then give some background on general diffusion models.
\subsection{The 2D XY model}
The 2D XY model is defined as spins on a two dimensional $L\times L$ lattice with periodic boundary conditions. The spins are periodic between $0$ to $2\pi$. This model has both local and long range correlations between the spins which gives rise to interesting physics. It exhibits a Berezinskii-Kosterlitz-Thouless (BKT) phase transition at the critical temperature $T_{c} = 0.89$ \cite{bkt_1973, xy_critical_temp}. At high temperature ($T> T_{c}$), the correlation between the spins decay exponentially with the distance between the spins while the correlations decay as a power law at low temperature ($T< T_{c}$) \cite{bkt_1973}. These long-range correlations can make sampling challenging for traditional Monte Carlo methods and in fact, at low temperature, the thermalization time scales exponentially with lattice size \cite{glauber_mixing}. This makes the XY model a very interesting model to study. We denote the angles of an $L\times L$ 2D XY model lattice by $\phi\in [0,2\pi)^{L\times L}$ and the inverse temperature is defined as $\beta = 1/T$.
\subsection{Diffusion model}
Diffusion model is a generative model which learns how to produce samples from some data distribution progressively removing noise through a stochastic process \cite{ddpm, ncsn, anderson}. Diffusion models are trained using a score network which approximates the score of the data distribution across noise levels. Stochastic differential equations (SDEs) have been used to model the noising and denoising process in diffusion models \cite{song-score-based}. 

Traditionally, such noising and denoising processes were defined on Euclidean spaces, but recent work has shown that such diffusion processes can be defined on compact Riemannian Manifolds \cite{riemanian-diffusion}. Samples from a $L\times L$ XY model lie on a $L^2$ dimensional hypertorus (denoted by $\mathbb{T}^{L^2}$), which is an example of a compact Riemannian Manifold. The first torsional diffusion framework was introduced in \cite{torsional} where the authors implemented an equivariant diffusion model on $\mathbb{T}^d$ for molecule conformer generation where $d$ refers to the number of torsion angles. In this work we use the standard score-based diffusion model with a rotationally invariant score network to generate samples for the XY model across different sizes and temperatures. The forward noising process is defined by:
\begin{equation}\label{forward-equation}
    d\phi = f(\phi,t)dt + g(t)dw
\end{equation}
where $w$ is a Wiener process. $f(\phi,t)$ and $g(t)$ are functions which can be chosen. We use a Variance Exploding SDE and thus we choose $f(\phi,t) = 0$ and $g(t) = \sqrt{\frac{d}{dt}\sigma^{2}(t)}$ where $\sigma(t)$ is an increasing noise schedule. The forward process converges to a uniform distribution after sufficiently large number of steps instead of a Gaussian distribution due to the periodicity in the angles. This score-based diffusion framework uses the following reverse SDE below to generate samples from uniform noise:
\begin{equation}\label{reverse-equation}
    d\phi_{t} = [f(\phi_{t}, t) - g^{2}(t)\nabla_{\phi}\log{p(\phi_{t})}]dt+g(t)dw
\end{equation}
The diffusion model is trained using the score-matching loss to approximate the score $\nabla_{\phi}\log p(\phi_t)$ by learning a score network $s_\theta$, where $\theta$ denotes the network parameters \cite{song-score-based}. In our case, we train a temperature-conditioned score network to learn the temperature dependent score. 

\section{Method}\label{sec:method}

 \begin{figure}[h]
    \centering
        \centering
        \includegraphics[width=0.85\linewidth]{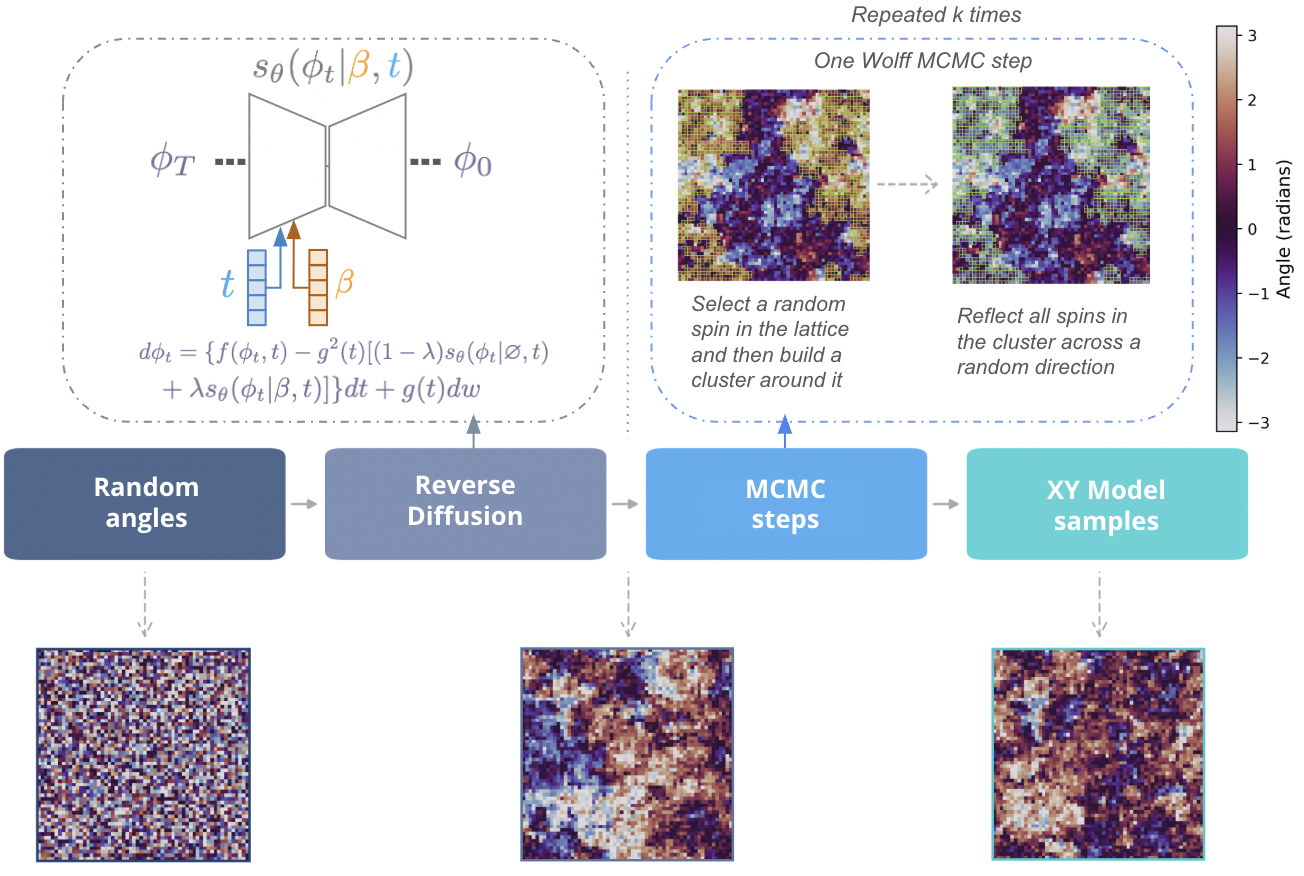}
        \caption{The figure shows the whole pipeline of diffusion-warmed MCMC. Random configurations are passed through the reverse diffusion process at a given temperature. The colourbar represents the spin values between $-\pi$ to $\pi$. The reverse diffusion process applies the stochastic differential equation sequentially from $t = 1$ to $t = 0$ as shown in the top left panel. $k$ steps of Wolff MCMC are then applied on diffusion-warmed samples. The top right panel shows one step of Wolff MCMC where a cluster is first formed and then all the spins in the cluster are reflected across a random direction. The highlighted spins in the image show the spins belonging in the cluster. The last row demonstrates the full pipeline at $\beta = 1$ for a $60\times 60$ system. Here reverse diffusion is done for $500$ steps using the model trained at $20\times20$ and $k$ is chosen to be $30$ for Wolff MCMC.}
        \label{fig:method}
\end{figure}

In this paper we introduce a diffusion-warmed MCMC method to generate Boltzmann samples from the 2D XY model. Our method first involves training a rotationally invariant diffusion model on a fixed small lattice size conditioned at a wide range of inverse temperatures. The trained diffusion model is then used to generate samples of bigger lattice sizes followed by a few Wolff MCMC steps. Wolff MCMC is required as the diffusion samples do not have the right thermodynamic properties as discussed in \cref{sec:results}. This implies that the diffusion samples at larger lattice sizes are a very good warm start  for Wolff MCMC resulting in significantly faster thermalization to the Boltzmann distribution compared to standard Wolff MCMC starting from random initialized configurations. In the following section, we first introduce the diffusion method and then explain how we extrapolate to larger lattices.

\subsection{Temperature-conditional diffusion model}\label{sec:diffusion}
We use a temperature-conditioned score network that is invariant to global rotations and compatible with periodic boundary conditions. In the conditional generation setting, the goal is to estimate the score associated with the conditional reverse time SDE. The reverse time SDE with temperature conditioning is: 
\begin{equation}\label{conditional-reverse-equation}
    d\phi_{t} = [f(\phi_{t}, t) - g^{2}(t)\nabla_{\phi}\log{p(\phi_{t}|\beta)}]dt+g(t)dw
\end{equation}
Several methods exist to train such a conditional model, but to the best of our knowledge, the state of the art technique is Classifier-Free Guidance (CFG) \cite{cfg}. We thus adopt this CFG framework which involves training a single model jointly with the conditional and unconditional score. At the inference stage, both the conditional and unconditional scores are used to guide the reverse diffusion process. 
\paragraph{Training} We train the score network with the score matching loss using the CFG training strategy. To jointly train the network with both the conditional and unconditional score, we apply a mask to probabilistically remove some of the conditioning labels ($\beta$). We observed that dropping the conditioning labels with probability $10\%$ gives us the best results. We trained models using $20\times20$ XY model samples across a grid of $11$ inverse temperatures from $\beta = 0.5$ to $\beta = 1.6$. 
\paragraph{Score network} In this work, since we are primarily interested in size extrapolation, the score network is composed of CNNs. We use periodic convolutions because the 2D XY model has periodic boundary conditions as described in \cref{sec:background}. Furthermore the input to the network are angles and thus we have a sine-cosine embedding in the network to handle the periodicity. Global rotational invariance is ensured by preprocessing the data using global rotations such that all the samples have mean $0$ during training.
\paragraph{Inference} At the inference stage we use the CFG based sampling method with both the conditional and the unconditional scores. The reverse time SDE that used for sampling is given by:
\begin{equation}\label{conditional--final-reverse-equation}
    d\phi_{t} = \{f(\phi_{t}, t) - g^{2}(t)[(1-\lambda)s_{\theta}(\phi_{t}|\varnothing,t) + \lambda s_{\theta}(\phi_{t}|\beta,t)]\}dt+g(t)dw
\end{equation}
Here $s_{\theta}(\phi_{t}|\varnothing,t)$ corresponds to the unconditional score predicted by the neural network, $s_{\theta}(\phi_{t}|\beta,t)$ is the conditional score, and $\lambda$ is the guidance scale defining the strength of the conditional score. Stronger guidance (higher $\lambda$) produces samples which adhere more closely to the given conditioning ($\beta$), but at a cost of lower variance. In physics systems like the XY model, the spins become more aligned as temperature decreases. Accordingly, we use an annealed guidance schedule where the guidance scale is progressively increased with $\beta$. After generating the samples, we perform another random global rotation to each sample to recover the global $U(1)$ symmetry of XY model.
\subsection{Generalization in lattice size}
The trained model can be used to generate samples on larger lattices by initializing the reverse diffusion process with uniformly random angles on an $L\times L$ grid for some $L$ beyond the size trained on. However, we show in \cref{sec:results} that doing inference for an arbitrary $L\times L$ system using a model trained on $20\times 20$ systems does not produce the correct samples. A model trained on a $20\times 20$ lattice with periodic boundary conditions captures artificial correlations between spins separated by more than $10$ lattice sites. Furthermore, the XY model has long wavelength spin-waves at low temperature \cite{spin-waves}. Models trained on small lattices will not have such long wavelength spin waves as the lattice size limits the maximum wavelength. This is why this model trained on $20\times20$ systems will not be able to faithfully produce long wavelength spin waves at low temperature for bigger lattices. Hence, a naive extrapolation in size during inference will fail to capture global thermodynamic properties. However, we show that enhancing the diffusion samples with a few Wolff MCMC steps produces samples which capture the right physics of the XY model. 

Wolff MCMC involves forming a cluster of spins based on the temperature and then flipping all the spins in the cluster \cite{wolff, wolff_good_explanation}. We describe the Wolff MCMC method in detail in \cref{app:wolff}. Applying a few Wolff MCMC steps to samples obtained from the diffusion model allows us to obtain high quality XY model samples at arbitrary lattice sizes. In the following section, we provide empirical results to compare diffusion-warmed MCMC against traditional MCMC for the XY model.


\section{Results}\label{sec:results}
The main focus of this work is size extrapolation. We demonstrate that our approach outperforms traditional Wolff MCMC in thermalization time to the Boltzmann distribution. This shows that the diffusion model effectively produces a very good initialization (warm start) for Wolff MCMC. First we discuss some results which illustrate why the diffusion model does not produce the right observables when we extrapolate in size. Then we discuss how subsequent Wolff MCMC updates correct the observables.

\subsection{Correlation function}\label{sec:results-correlation}
The spin-spin correlation function $C(r)$ measures the correlation between spins separated by a distance $r$. This is given by:
\begin{equation}
C(r) = \braket{s_{i}\cdot s_{i+r}} = \braket{\cos(\phi_{i})\cos(\phi_{i+r})+\sin(\phi_{i})\sin(\phi_{i+r})} = \braket{\cos(\phi_{i} - \phi_{i+r})}
\end{equation}
We observe that there is a discrepancy in the correlations in the samples produced by the diffusion model versus ground truth correlations from long MCMC runs.

\begin{figure}[t]
    \centering
        \centering
        \includegraphics[width=0.7\linewidth]{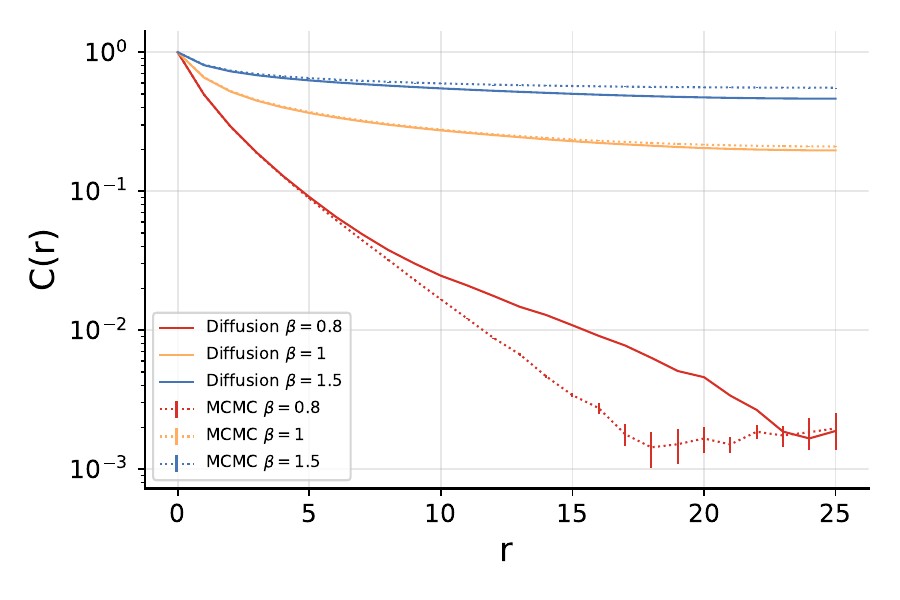}
        \caption{The figure shows spin-spin correlation $C(r)$ for samples produced by diffusion model and MCMC against $r$ at three different temperatures. 
        The diffusion model has been trained on $20\times20$ lattice and then used to generate samples for $50\times50$ lattice at different temperatures. The vertical lines for MCMC represent standard deviation error bars around the mean.}
        \label{fig:correlations}
\end{figure}
In \cref{fig:correlations}, we compare the spin-spin correlations of samples produced by the diffusion model and MCMC. The $C(r)$ captured by the diffusion model becomes less accurate with decreasing temperature. At high temperature (i.e. above the critical temperature),  the $C(r)$ produced by the diffusion model seems reasonably good, but they still do not match perfectly. However it is interesting to note that the model seems to under estimate correlations at low temperature ($\beta = 1.5$) and over estimate correlations at high temperature ($\beta = 0.8$). 

\begin{figure}[b]
    \centering
        \centering
        \includegraphics[width=\linewidth]{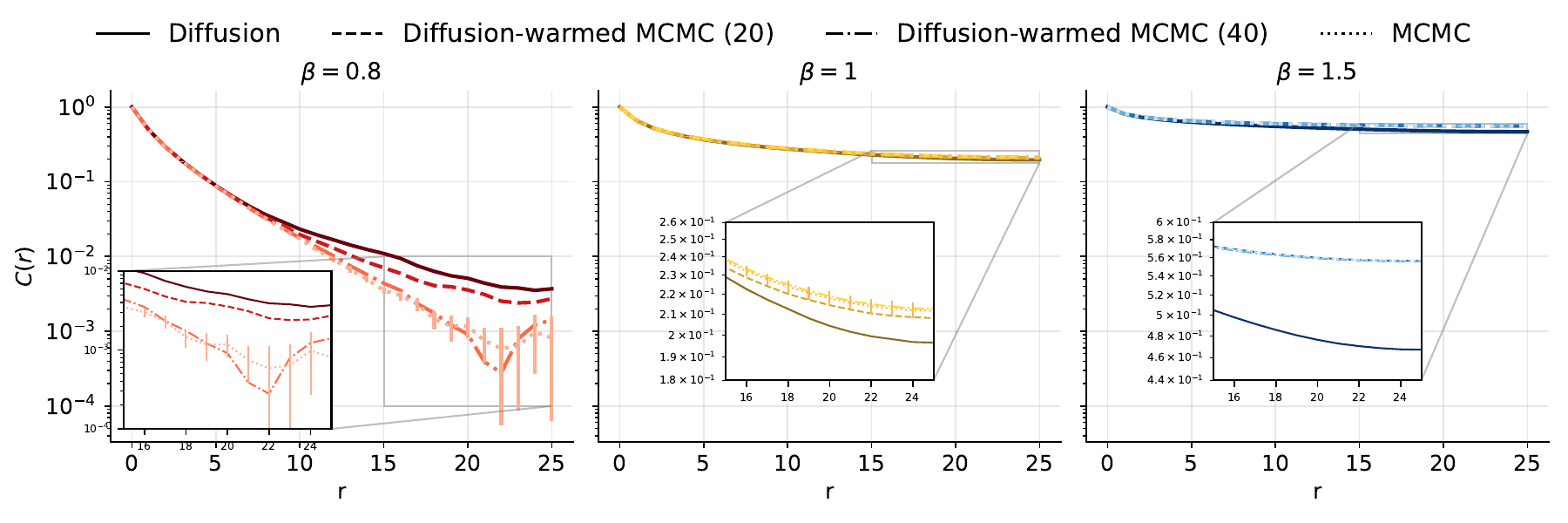}
        \caption{The figure demonstrates that a small number of Wolff MCMC updates is sufficient to correct the spin-spin correlations of samples generated by the diffusion model for a $50\times 50$ system. The lines labeled diffusion-warmed MCMC ($20$) and diffusion-warmed MCMC ($40$) represent the spin-spin correlation function $C(r)$ computed from samples generated by the diffusion model followed by $20$ and $40$ Wolff MCMC updates respectively. The error bars represent the standard deviation around the mean. The dotted line represents the $C(r)$ obtained from fully thermalized systems obtained from long MCMC runs. The inset plots show the long range correlations in more detail.}
        \label{fig:correlations-correction}
\end{figure}
This mismatch is expected, as we have only trained the model on $20\times 20$ lattice and thus it cannot be expected to generalize perfectly for a larger lattice. Therefore, we use Wolff MCMC steps to fix the spin-spin correlations. In \cref{fig:correlations-correction} we use the diffusion model trained on $20\times20$ system to generate $1000$ samples of $50\times 50$ system at $\beta = 1$. We then apply Wolff updates to each of the $1000$ samples and find that after approximately $30$ steps, the corrected $C(r)$ matches that of $1000$ fully thermalized MCMC samples.

\subsection{Scaling analysis}

In this section, we look at the scaling of thermalization time using Wolff MCMC versus diffusion-warmed MCMC. Determining whether a system has thermalized is an exponentially hard problem \cite{thermalization-complexity} and different observables may converge at different rates \cite{fast_thermalization, thermalization_quantum}. In practice, thermalization time is estimated using long MCMC runs by monitoring convergence of all observables. While convergence of observables does not imply that a system has reached equilibrium \cite{thermalization-time-scales}, this is the best that can be done for such physics systems as there is no clear metric showing a system has thermalized \cite{thermalization_molecule, thermalization_quantum, helicity-numerical}. Hence, we obtain ground truth data of all observables using long MCMC runs at all the temperatures and the lattice sizes given in \cref{fig:main-result}.

Different observables will converge at different times and thus we define the thermalization time to be the time taken for the slowest observables to converge \cite{thermalization-time-scales}. We tracked expected energy, magnetization, helicity modulus and heat capacity for both diffusion-warmed MCMC and MCMC starting from $1000$ random spin configurations. More details on how we concretely compute this thermalization time are given in \cref{app:thermalization}.

Sampling from spin systems is expected to take exponentially long thermalization time with respect to system size $L$ at low temperature \cite{glauber_mixing}. We observe in \cref{fig:main-result} (right panel) that both the methods seem to require exponential time for thermalization with respect to $L$ at $\beta = 1.5$. At higher temperatures, we expect thermalization time to scale polynomially with system size with MCMC \cite{wolff, wolff_good_explanation, glauber_mixing}. The diffusion model run time is expected to scale polynomially with system size as the score network is composed of CNNs, which scale polynomially with $L$ \cite{deep_learning_time_complexity}. However, the exact scaling of sampling using diffusion and then doing MCMC is not obvious from \cref{fig:main-result} and requires further studies.
Nevertheless, the results in \cref{fig:main-result} illustrate that diffusion-warmed MCMC reaches equilibrium faster than the MCMC method by around an order of magnitude for all the system sizes and temperatures we have studied. 
Improvements in thermalization time is very important for Gibbs sampling in general, especially at low temperature, where sampling is known to be hard due to metastability and high energy barriers \cite{glauber_mixing, stat-mech-book}. \cref{fig:main-result} shows that at $\beta = 1.5$ the thermalization time is approximately $10$ times faster for the diffusion-warmed MCMC compared to Wolff MCMC.
\subsection{Observables}
\begin{figure}[h]
    \centering
        \centering
        \includegraphics[width=1\linewidth]{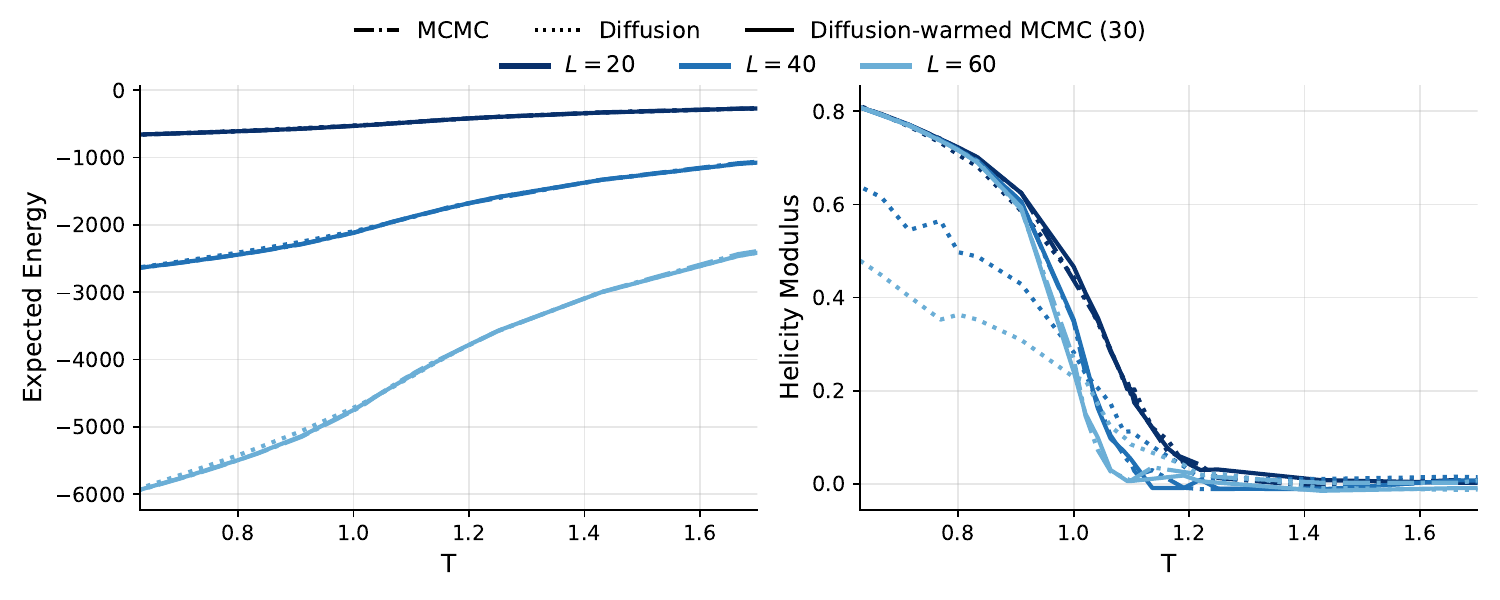}
        \caption{The figure shows the expected energy and helicity modulus across a wide range of temperatures, demonstrating that diffusion-warmed MCMC with only $30$ Wolff updates is sufficient to accurately reproduce equilibrium observables for different system sizes. The data in the lines labeled MCMC were obtained from with long MCMC runs and the lines labeled diffusion-warmed MCMC ($30$) represent the spin-spin correlation function $C(r)$ computed from samples generated by the diffusion model followed by $30$ Wolff MCMC updates. The solid lines were obtained by applying $30$ Wolff MCMC steps on the samples produced by the diffusion model.}
        \label{fig:energy-helicity-comparison}
\end{figure}

\paragraph{Helicity Modulus}
Helicity modulus measures the stiffness in XY model. It is a global observable and thus it is very sensitive to long range correlations unlike energy and magnetization, which are local observables. Intuitively it is a measure of resistance against a twist in the system in a particular direction \cite{helicity-numerical, sandvik-helicity}. This quantity undergoes a sharp transition at the critical temperature in the thermodynamic limit \cite{bkt_1973}. Thus we expect to see this transition to get sharper as we increase system size. In our computational results we used $\rho_{s} = \frac{((\rho_s)_{x})+(\rho_s)_{y})}{2}$ to get the results in \cref{fig:energy-helicity-comparison}. The helicity modulus along a dimension $a$ for $N$ samples at inverse temperature $\beta$ is given by:
\begin{equation}\label{helicity-equation}
(\rho_s)_{a} = \frac{1}{N} \left\langle \sum_{\langle i,j \rangle_a} \cos(\phi_i - \phi_j) \right\rangle 
- \frac{\beta}{N} \left\langle \left[ \sum_{\langle i,j \rangle_a} \sin(\phi_i - \phi_j) \right]^2 \right\rangle
\end{equation}

In \cref{fig:energy-helicity-comparison}, the model was trained on $20\times20$ samples and thus we expect the helicity modulus produced by the diffusion samples to match the ground truth observables for $L = 20$. The figure shows that the diffusion model samples accurately capture the trend for the $20\times 20$ system. However, for larger systems, the diffusion model samples underestimate the helicity modulus at low temperature. Recall that in \cref{fig:correlations} we observed the diffusion model underestimates the spin-spin correlations at low temperatures (i.e. below critical temperature) implying that the samples are less aligned than they should be. Hence, as the alignment of the spins is underestimated (due to underestimation in $C(r)$), the first term $\frac{1}{N} \left\langle \sum_{\langle i,j \rangle_a} \cos(\phi_i - \phi_j) \right\rangle$ is underestimated and the second term $\frac{\beta}{N} \left\langle \left[ \sum_{\langle i,j \rangle_a} \sin(\phi_i - \phi_j) \right]^2 \right\rangle$ is overestimated. This results in this underestimation of helicity modulus at low temperature. Nevertheless, as shown in the figure, we can fix the error in helicity modulus after $30$ steps of Wolff MCMC on the diffusion samples. 
\paragraph{Expected Energy}
The energy of the nearest neighbour XY model is given by:
\begin{equation}
E_{XY} =  \sum_{\langle i,j \rangle} \cos(\phi_i - \phi_j)
\end{equation}
In \cref{fig:energy-helicity-comparison} we show the expected energy, $\braket{E_{XY}}$, against temperature using $1000$ samples. The energies of samples produced by diffusion and MCMC match quite well at $L = 20$ as expected as we trained on $L = 20$. The diffusion energies are also very close to the ground truth MCMC energies at high temperature for bigger sizes. 

At low temperature we can see that the energies are close, but do not match exactly. This can also be explained using the correlation plots in \cref{fig:correlations}. Note that at low temperature, in \cref{fig:correlations},  the diffusion correlations are slightly lower than the MCMC correlations at small $r$. This underestimation of the correlations at small $r$ causes the energy to be slightly higher than the actual energy at low temperatures which we can observe in \cref{fig:energy-helicity-comparison} at low temperature for $L = 40$ and $L = 60$. Nevertheless, the energy converges to the ground truth in $30$ Wolff MCMC steps.
\section{Conclusion}\label{sec:conclusion}
In this work we introduced a Gibbs sampling method for the XY model based on a diffusion model that allows reliable generation of configurations for lattice sizes beyond the training regime. We demonstrate that this method provides a very good warm start for MCMC which subsequently reduces the total thermalization time by roughly an order of magnitude as compared to traditional MCMC starting from random configurations. While this work clearly demonstrates a significant reduction in thermalization time, the scaling with respect to system size requires further studies. Additionally, it will be interesting to test this method for other systems beyond the XY model and extrapolation of other physical quantities like temperature. We believe this work highlights the potential of such generative models to facilitate large scale accurate simulation of condensed matter systems. This has implications in condensed matter and computational physics where scaling to large system sizes is essential, particularly for reliably detecting phase transitions in the thermodynamic limit.

\section*{Acknowledgement}

Authors thank Vichayuth~Imchitr, Tarun~Kumar, Shaeer~Moeed, Schuyler~Moss, Abhi~Sarma, and Yi-Hong~Teoh for useful technical conversations. This work is supported by NSERC Discovery grant RGPIN-2022-03339, and the Quantum Computing Challenge Program AQC-206 at the National Research Council of Canada (NRC). Authors further acknowledge the support of Perimeter Institute for Theoretical Physics, supported in part by the Government of Canada through the Department of Innovation, Science and Economic Development Canada (ISED), and by the Province of Ontario through the Ministry of Economic Development, Job Creation, and Trade.

\bibliographystyle{plainurl}
\bibliography{bibliography}


\appendix

\section{Finding thermalization time}
\label{app:thermalization}
In this section we describe how we computed the thermalization time. As already mentioned in \ref{sec:results}, there is no standard way to compute this thermalization time for condensed matter systems as different observables thermalize at different times in general and convergence of some observable does not imply the whole system has thermalized. We first run long MCMC simulations for all system sizes and temperatures considered. Let $\widetilde{\mathcal{O}}$ denote the value of an observable $\mathcal{O}$ obtained from the long MCMC run after thermalization. Now we define $\tau_{eq}(\mathcal{O,\epsilon_{\mathcal{O}}})$ as number of MCMC steps required for observable $\mathcal{O}$ to converge to $\widetilde{\mathcal{O}}$ with relative error tolerance $\epsilon_{\mathcal{O}}$. Note that the error tolerance $\epsilon_{\mathcal{O}}$ depends on the observable. $\mathcal{O}(\tau)$ is used to denote the value of the observable as a function of MCMC steps, $\tau$. We define $\tau_{eq}(\mathcal{O}, \epsilon_{\mathcal{O}})$ as:
\begin{equation}
\tau_{\mathrm{eq}}(\mathcal{O},\epsilon_{\mathcal{O}})
:= \min \left\{ \tau \;:\;
\left|\frac{\mathcal{O}(\tau) - \widetilde{\mathcal{O}}}{\widetilde{\mathcal{O}}}\right| \le \epsilon_{\mathcal{O}}
\right\}
\end{equation}
We track energy ($E$), magnetization ($M$), helicity modulus ($\rho_{s}$) and heat capacity $C_{v}$ to compute the thermalization time. Note that $E \text{ and } M$ are local observables and $\rho_{s} \text{ and }C_{v}$ are global observables. We set $\epsilon_{\mathcal{O}} = 0.05$ for $E$, $M$ and $\rho_{s}$, but we allow a bigger relative error tolerance of $\epsilon_{\mathcal{O}} = 0.1$ for $C_{v}$. This is because the heat capacity is a  variance term and thus the variance is $C_{v}$ is expected to be large for small number of samples. In our analysis, we generated $1000$ samples using both the pure MCMC and diffusion based method due to computational resource limitations and this is why we require a higher tolerance for $C_{v}$. Then, after we obtain the $\tau_{eq}(\mathcal{O,\epsilon_{\mathcal{O}}})$ for each of the observables, we say that the number of MCMC steps required for thermalization is:
\begin{equation}
    \tau_{eq} = \max\left\{\tau_{eq}(\mathcal{O,\epsilon_{\mathcal{O}}})\;:\; \mathcal{O}\in\{E,M,\rho_{s},C_{v}\}\right\}
\end{equation}
For MCMC, thermalization time is the wall clock time for $\tau_{eq}$ Wolff MCMC steps. Additionally, for diffusion-warmed MCMC, thermalization time is the total time taken for reverse diffusion and wall clock time for $\tau_{eq}$ Wolff MCMC steps.
\section{Wolff MCMC}\label{app:wolff}
Wolff MCMC is a form of cluster update where a cluster is first probabilistically built and then all the spins in the cluster are updated simultaneously. We define cluster as a set of lattice sites such that all the lattice sites are connected. We denote a cluster as $C$ where $C$ contains the list of the lattice sites. We denote each lattice site as $(i,j)$. 
A single Wolff MCMC step involves the following steps \cite{wolff, Lauren_thesis}:

\begin{enumerate}
    \item Initialize an empty cluster $C$, an empty queue $Q$ and an empty list $M$ containing sites which have been visited
    \item Choose a random site $(k,l)$ in the lattice and add the site $(k,l)$ to the cluster $C$, the queue $Q$ and list $M$.
    \item choose a random $2$ dimensional unit vector $v$
    \item while $Q$ is not empty
    \begin{enumerate}
        \item remove a spin $s$ from $Q$  
        \item for each neighbour $n$ of $s$, if $n$ is not present in $M$ then add $n$ to $C$, $Q$ and $M$ with probability
        $$p_{add} = 1-\exp{(\min{(0, -2\beta (s\cdot v)(n\cdot v))})}$$
    \end{enumerate}
\end{enumerate}
The cluster is formed when the queue $Q$ is empty. Now we reflect all the spins in the cluster. Let's say the set of all spins in the cluster is denoted by $S_{c}$. These spins are reflected about the $v$ direction and thus, the spins are modified as:
\begin{equation*}
    s\rightarrow s - 2 (v\cdot s)v \quad \forall s\in S_{c}
\end{equation*}
This constitutes a single Wolff MCMC step, and the procedure is repeated for subsequent updates.

\section{Comparing diffusion-warmed MCMC with random tiling}\label{app:tiling}
\begin{figure}[h]
    \centering
        \centering        \includegraphics[width=0.7\linewidth]{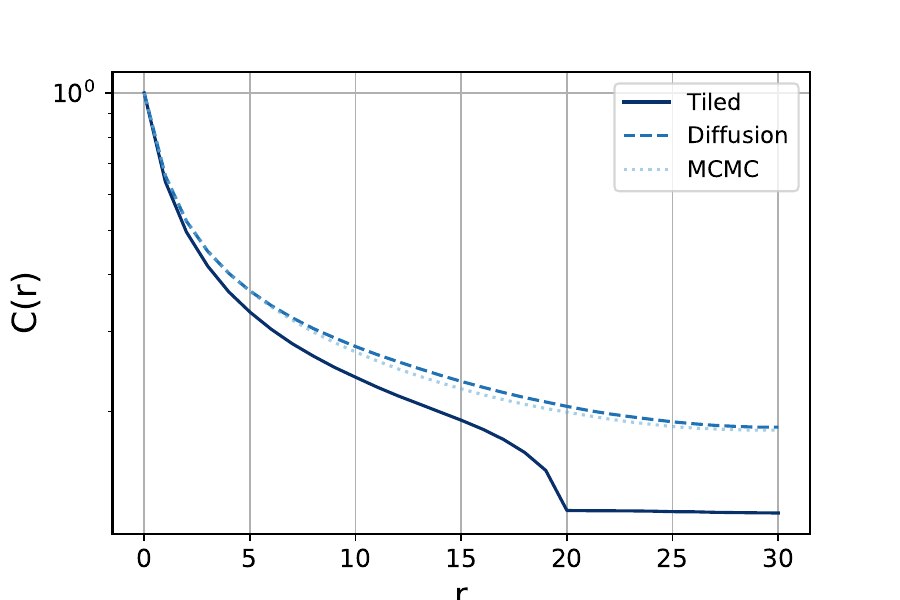}
        \caption{The figure shows how $C(r)$ for diffusion samples (without any MCMC step) compares against randomly tiled $20\times20$ samples for a $60\times 60$ lattice at $\beta = 1$. The $20\times20$ samples were also at $\beta = 1$.}
        \label{fig:tiled}
\end{figure}

Starting an MCMC chain from a random initialization generally requires longer to thermalize than starting a chain from a better initialization. Our experiments have revealed that starting a chain from a tiled configuration significantly reduces thermalization time than starting from random configuration. Tiled configuration refers to configurations built from thermalized samples from smaller lattice. In our example in \cref{fig:tiled} we take random $20\times 20$ MCMC samples at $\beta = 1$ and tile them together to make a $60\times 60$ configuration. Note that we perform global rotations on each tile to ensure that each of the tiles have the same mean spin direction and then perform another random global rotation on the $60\times$ tiled configuration to restore the $U(1)$ symmetry. We considered tiling $20\times 20$ samples to ensure compatibility with the diffusion model samples as our diffusion model has been trained on $20\times 20$. This result reveals that the diffusion samples indeed captures better correlations than the tiled samples. The tile samples seem to have artificial artifacts closer to $r = 20$ and also significantly underestimate the correlations. This shows that generating samples using the diffusion model is indeed better than just tiling smaller samples together.

\section{Annealed classifier-free guidance}
In this section we compare classifier-Free guidance results with and without annealing guidance scale. Note that the guidance scale is only relevant during inference and not during training. Experiments revealed that CFG most significantly affects the heat capacity observable. This makes sense as heat capacity is related to variance in energy and we know that the guidance scale directly affects the variance of the samples \cite{cfg}. 

\begin{figure}[h]
    \centering
        \centering        \includegraphics[width=0.7\linewidth]{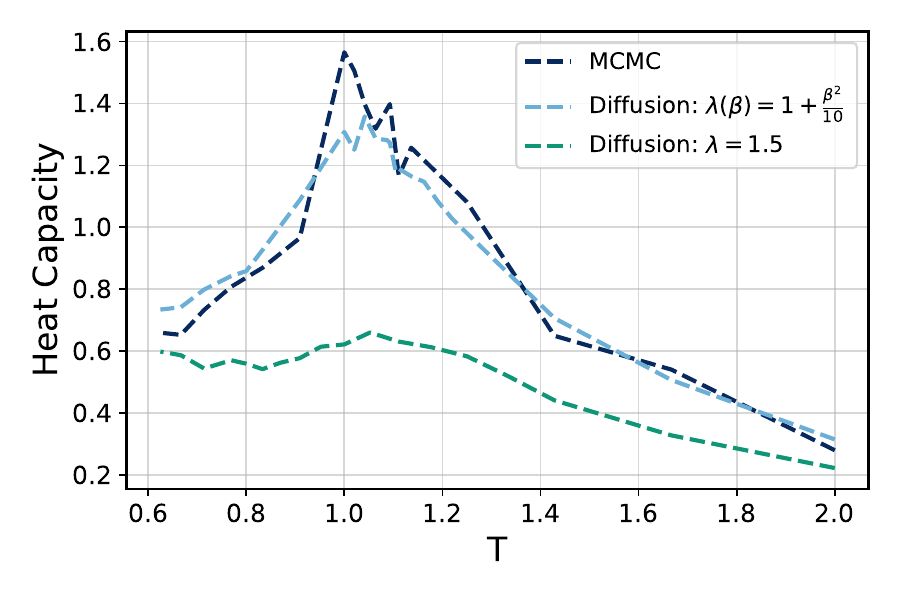}
        \caption{The figure shows how heat capacity changes with different annealing strategies during inference.}
        \label{fig:heat-capacity-anneal-comparison}
\end{figure}

We found heuristically that the anneal schedule for guidance scale $\lambda$ is $\lambda(\beta) = 1 + \frac{\beta^2}{10}$. In \cref{fig:heat-capacity-anneal-comparison} we compare the performance of the annealed guidance with a fixed guidance scale of $1.5$. Note that no Wolff steps have been done on the diffusion samples here and we can see that the heat capacity of the diffusion samples produced by the annealed CFG method are clearly better than the samples produced by the fixed guidance scale.
\section{Experimental setup}\label{app:training-setup}
In this section we discuss training data generation and the setup of the experiments. All experiments were run on a Google Cloud Platform (GCP) instance (n1-standard-4 machine type with 4 vCPUs and 15 GB RAM) using a single NVIDIA Tesla T4 GPU and 32 GB standard persistent boot disk.
\paragraph{Training data generation}
The training data has been generated using Wolff MCMC \cite{wolff} which was introduced in \cref{app:wolff}. Wolff MCMC updates the spin configurations by forming clusters where the cluster size depends on the temperature and then collectively flipping all the spins in the cluster. For each configuration ($L$, $\beta$) a single chain was used to generate the data. We used $10,000$ Wolff steps for thermalization and then collected the samples after every $10$ steps to ensure the samples are not correlated \cite{autocorrelation, wolff_good_explanation}. We generate $20,000$ samples for each $\beta$ in $\beta\in [0.5,1.6]$ with increments for $0.1$ for $L = 20$ to obtain training data. $70\%$ of the data was used for training the model while the rest were used for validation.

\paragraph{Training details}
The model has been trained using the score-based diffusion framework \cite{song-score-based} with Classifier-free Guidance \cite{cfg} jointly at all the temperatures as we outlined in \cref{sec:diffusion}. The score network had $510$K parameters and we trained using a single Tesla T4 GPU for $1150$ epoch. Note that the $510$K parameters also includes the parameters coming from the time embedding network and the $\beta$ conditioning network. We used periodic convolutions in the score network as our score network had periodic boundary conditions. The input and output convolution layers had kernels of size $5\times 5$ to capture more correlations than just nearest neighbour correlations. 

\begin{figure}[h]
\centering

\begin{minipage}{0.42\linewidth}
\centering
\captionof{table}{Training hyperparameters and score network details}
\label{tab:hyperparameters}
\begin{tabular}{ll}
\toprule
\textbf{Hyperparameter} & \textbf{Value/Specification} \\
\midrule
Learning rate           & $1 \times 10^{-4}$ \\
Batch size              & $500$ \\
Optimizer               & Adam \cite{adam} \\
$\sigma_{\min}$         & $0.01$ \\
$\sigma_{\max}$         & $50$ \\
Input encoding & $(\cos\theta, \sin\theta)$ \\
Padding & Circular \\
Residual blocks & $3$ \\
Time embedding & Sinusoidal \cite{song-score-based}\\

\bottomrule
\end{tabular}
\end{minipage}
\hfill
\begin{minipage}{0.56\linewidth}
\centering
\includegraphics[width=\linewidth]{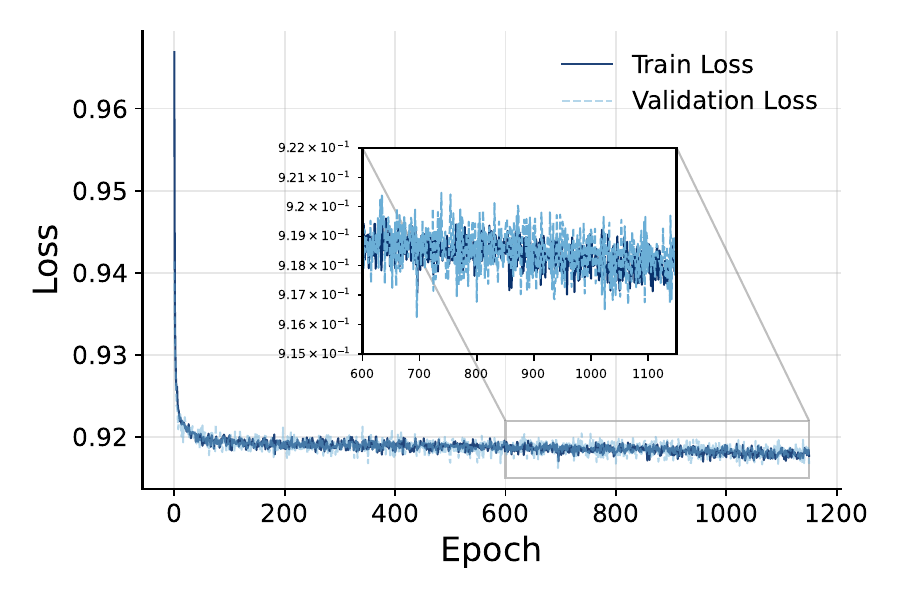}
\caption{Training and validation loss curves.}
\label{fig:loss_plot}
\end{minipage}

\end{figure}
We have done some hyperparameter tuning in the model architecture and noise schedule,  but we found that hyperparameter tuning does not drastically impact the results. The primary focus of this work is to show that a diffusion model trained on small size ($L = 20$) can be used to prepare a very good warm start for a larger lattice. The model trained with configurations used in table \ref{tab:hyperparameters} indeed shows such results as we demonstrated in \cref{sec:results}. However, we acknowledge there are several ways to improve the model by making the network stronger or by improving the noise schedule. For example, we observed that a lower $\sigma_{max}$ helps the model to obtain a better loss which consequently produces more accurate correlations when doing inference on bigger sizes. However, the correlations still do not match as a model trained on small sizes do not see the long wavelength structures which are present in bigger lattices, as we discussed in \cref{sec:results-correlation}. Hence, Wolff MCMC steps are still required. 
 
\section{Visualization of XY samples}
\begin{figure}[h]
    \centering
        \centering        \includegraphics[width=0.9\linewidth]{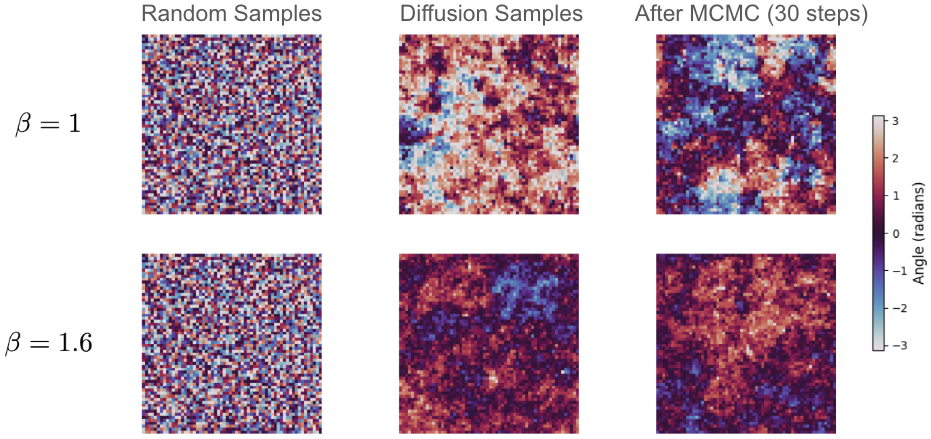}
        \caption{The figure shows samples produced after diffusion and then after $30$ MCMC steps. The angles are represented by the colour bar on the right.}
        \label{fig:visualize}
\end{figure}

In this section we visualize some of the samples generated at different temperatures. The diffusion samples in \cref{fig:visualize} were obtained from a $60\times 60$ lattice using a model trained on $20\times 20$. Note that there is a clear difference in the quality of the samples between $\beta = 1$ and $\beta = 1.6$. The diffusion samples for $\beta = 1.6$ indeed look more correlated compared to the corresponding samples for $\beta = 1$. This is expected as the correlations between spins increases as we decrease the temperature.

\section{Score network architecture}\label{app:architecture}
Here we show the architecture of the score network.
\begin{figure}[htbp]
    \centering
        \centering        \includegraphics[width=0.85\linewidth]{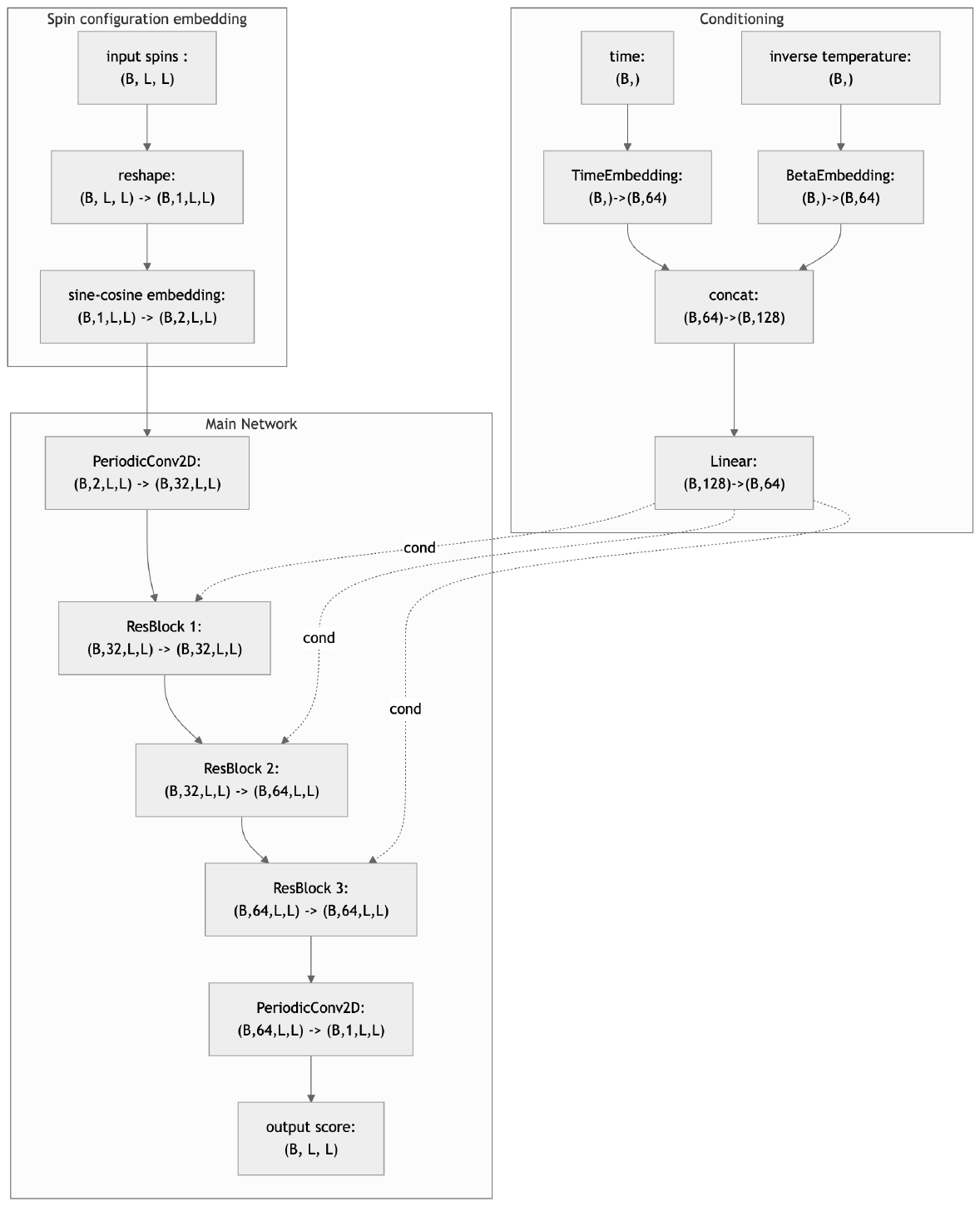}
        \caption{The figure shows the architecture of the score network. The input spins are used to first get a sine and cosine embedding. The conditioning for both time and inverse temperature $\beta$ is done using sinusoidal time embedding \cite{song-score-based}. Periodic convolutions are used in the main body of the network. The residual networks are also composed of convolution layers and the embeddings produced from the conditioning are used in the network through Feature-wise Linear Modulation (FiLM) \cite{film}. The final output is the estimated score produced by the network. The network has in total $510$K trainable parameters.}
        \label{fig:model-visualize}
\end{figure}


\end{document}